**Tackling cyclicity in causal models with cross-sectional data using a partial least square approach. Implication for the sequential model on internet appropriation.**

Lamberti, Giuseppe,[a*] Lopez-Sintas, Jordi,[b] Pandolfo, Giuseppe[c]

[a,c] *Department of Economics and Statistics, University of Naples Federico II, Naples, Italy*

[b] *Department of Business, Universitat Autònoma de Barcelona, Barcelona, Spain*

***Corresponding author**: [a] Giuseppe Lamberti, Department of Economics and Statistics, Room C1, University of Naples Federico II, Via Cintia, 21, 80126 Naples, Italy. Email: giuseppe.lamberti@unina.it, ORCID: http://orcid.org/0000-0002-8666-796X

[b] Jordi Lopez Sintas, Department of Business, School of Economics and Business, B Building, Room B1/118. Universitat Autònoma de Barcelona. Campus UAB 08193, Bellaterra (Barcelona) Spain. Email: jordi.lopez@uab.cat ORCID: http://orcid.org/0000-0001-5441-4039

[c] Giuseppe Pandolfo, Department of Economics and Statistics, Room C5, University of Naples Federico II, Via Cintia, 21, 80126 Naples, Italy. Email: giuseppe.pandalfo@unina.it ORCID: http://orcid.org/0000-0003-3668-0389

**Abstract**

Working with SEM and cross-sectional data, and depending on the studied phenomenon, assuming an acyclic model may mean that we obtain only a partial view of the mechanisms that explain causal relationships between a set of theoretical constructs, given that variables are treated as antecedents and consequences. Our two-step approach allows researchers to identify and measure cyclic effects when working with cross-sectional data and a PLS modelling algorithm. Using the resources and appropriation theory and the sequential model of internet appropriation, we demonstrate the importance of considering cyclic effects. Our results show that opportunities for physical access followed by digital skills acquisition enhance internet usage (acyclic effects), but also that internet usage intensity, in reverse, reinforces both digital skills and physical access (cyclic effects), supporting Norris' (2001) social stratification hypothesis regarding future evolution of the digital divide.

# 1. Introduction

Structural equation modelling (SEM) is a widely used multivariate technique for analysing complex systems of variables. The two most important prerequisites for SEM are (1) that a cause-effect relationship is hypothesized between variables, and (2) that most variables are not directly observable (i.e., they are latent) and so must be measured using a specific set of indicators.

The literature documents different algorithms that can be used to estimate these models. The classic approach is the well-known common factor-based (i.e., covariance-based) technique (Ullman and Bentler 2012), which considers latent variables to be common factors that explain co-variation between associated indicators (Sarstedt et al. 2016). This approach is generally based on strict assumptions regarding distributions and sample size. Partial least squares SEM (PLS-SEM) is an alternative approach (Hair et al. 2016; Wold 1985) that maximizes variations in the observed variables following a soft procedure that requires no assumptions regarding distributions and sample size (Hair et al. 2018).

In almost all cases, SEM application is conditioned by a critical hypothesis that Hyttinen (2012) defines as the "acyclic" hypothesis, which assumes that variables are antecedents or consequences, and that the causal relationship between variables can be estimated considering a specific time interval. In other words, the acyclic hypothesis operates in a SEM when a unidirectional sequential law regulates the relationship between variables. However, there are several situations in which variables influence each other in a cyclic manner, which means that we cannot assume acyclicity if we want to measure the complete structure of causal relationships between variables.

One sociology scenario which merits consideration in terms of acyclic and cyclic effects is how to explain social inequalities in internet adoption and use. The resources and appropriation (RA) theory (van Deursen and van Dijk 2015; van Dijk 2005, 2020) endeavours to explain how social inequalities are produced and reproduced in internet appropriation (the digital divide). Accordingly, the digital divide results from four constructs, namely, attitudes (i.e., reasons and motivations to use the internet), access (i.e., physical access to devices and connections), skills (i.e., mastery of the necessary technology), and usage (i.e., what use is made of the internet). The RA theory and model of internet appropriation operates on the assumption that these constructs influence each other sequentially, yet intuitively, it must be that internet usage also affect attitudes, access, and skills. Moreover, the model does not consider a time dimension, but reflects a snapshot of a single moment in time.

In proposing a new SEM algorithm that takes into account the time dimension, Asparouhov, Hamaker, and Muthén (2018) and Drton, Fox, and Wang (2019) resolved the problem of cyclicity in SEM; however, their proposal is based on longitudinal data, i.e., the same set of variables measured at different times, and longitudinal data are needed to apply their method. Since this kind of data is not usually available in SEM studies, analyses are cross-sectional.

In this article, we propose a two-step approach for measuring cyclic effects when working with cross-sectional data using a PLS algorithm. It is based on, first, estimating the classical causal model, and second, using the resulting dependent variables scores to estimate the cyclic effects.

Our approach applies the same philosophy as used for hierarchical models with higher-order constructs in PLS-SEM (Becker, Klein, and Wetzels 2012; Crocetta et al. 2021; Sarstedt et al. 2019). Hierarchical models use multidimensional constructs that exist at a higher level of abstraction but are quantified using lower-order sub-construct latent scores as proxies.

Summarizing, this article aims offering a methodological contribution in further developing PLS-SEM to measure a cyclic feedback loop. It also offers a sociological contribution in extending the sequential model of internet appropriation (van Deursen and van Dijk 2015; van Dijk 2005, 2020) to account for possible cyclic effects operating on the digital divide.

The rest of the paper is organized as follows. We first introduce the problem of cyclicity using an example from the sociology field. We next briefly describe the digital divide concept, the RA theory (van Dijk 2020), and the sequential model of internet appropriation (van Deursen and van Dijk 2015; van Dijk 2005, 2020), and also discuss the importance of considering cyclic effects in order to have a complete view of the social mechanisms that drive internet appropriation (Section 2). We then describe our two-step approach to estimating cyclic effects when using PLS-SEM (Section 3), and, using data from a Eurostat information and communication technology (ICT) usage survey, we illustrate how to estimate the cyclic effect in the model of internet appropriation (Section 4). Finally, we discuss the results and the methodological and substantive contributions of our research (Sections 5 and 6).

## 2. Conceptual background

### 2.1.1. RA theory and the digital divide

The digital divide is not just a metaphor, but points to a social problem: unequal access to and use of ICTs. Nor is the digital divide a theory, as underlined by van Dijk (2002), as it has successfully directed attention towards social, political, and academic debates (Attewell 2001). While the digital divide initially referred to unequal access to the ICTs, the evidence is that the concept is more complex than mere access (Norris 2001). Unequal access to ICTs nowadays is understood to be just a starting point for unequal endowment in devices, which also encompasses unequal attitudes, skills, and usage. However, the digital divide has several facets that need further explanation: (1) the theoretical relationship between unequal layers in the sequential process that culminates in unequal digital benefits (van Dijk 2002, 2005, 2012, 2020, van Deursen and van Dijk 2015), (2) the evolution of the digital divide (Norris 2001; van Dijk 2012), and (3) the unequal social distribution of the digital divide (van Deursen and van Dijk 2011, 2014, 2015).

The US National Telecommunications and Information Administration initially proposed digital divide as a term to refer to unequal penetration of computers among low-income groups, minorities, women, and the elderly (Norris 2001; Parsons and Hick 2008); this is now referred to as the first digital divide (Attewell 2001). Numerous policies have been introduced that aimed to close that access gap, yet subsequent research shows that a more persistent internet use gap emerged based on disparities in internet use, referred to as the second digital divide (Attewell 2001; Zillien and Hargittai 2009). In trying to understand the processes that generate the second divide, a third divide has been identified, related to the social and cultural advantages/disadvantages accruing to individuals with mor/fewer digital skills (Hargittai 2001; van Deursen and Helsper 2015). This skills gap suggests that to understand the digital divide better, we need to better understand digital divide levels and the relationships between them.

### 2.1.2. The sequential model of internet appropriation

The RA theory (van Dijk 2012; van Dijk 2005, 2020) proposes a sequential model of internet appropriation to explain the digital divide that sequentially links attitudes, physical access, digital skills, and usage. According to the RA theory, unequal benefits of internet use are grounded in unequal social distribution of those four constructs, briefly explained in relation to research findings below. For a more complete review, see Lamberti, Lopez-Sintas, and Sukphan (2021), van Dijk (2002, 2020), and van Deursen and van Dijk (2015).

*Attitude* (ATT). Attitude segments individuals according to those who are motivated and not motivated by the digital technologies. The evidence points to a positive attitude to digital technologies being fundamental to internet access and usage (Venkatesh et al. 2003), and to attitude also affecting the acquisition of digital skills (van Deursen and van Dijk 2015).

*Physical access* (PA). Physical access is the opportunity to access the internet (Dimaggio et al. 2004; Hargittai 2003), through devices such as desktops, laptops, tablets, smartphones, game consoles, and interactive television (van Deursen and van Dijk 2015, 2019) and through connectedness. Differences in equipment quality may affect what can be done while connected (Zillien and Hargittai 2009), while differences in physical access might evolve into different online skill levels (van Deursen and van Dijk 2010, 2019) and differing extents of internet usage (Kuhlemeier and Hemker 2007; Mossberger, Tolbert, and Hamilton 2012). The outcome of such differences could be the emergence of an underclass, reflecting a deepening of the digital divide (Napoli and Obar 2014).

*Digital skills* (DS). Digital skills reflect knowledge, capacity, and adeptness in using the internet efficiently and effectively (Hargittai 2002a; Hargittai and Shafer 2006). Digital skills are most widely used classified as medium-related skills, content-related skills, information skills, communication skills, safety skills, and problem-solving skills (Ferrari 2012; Hargittai, Piper, and Morris 2018; van Deursen and Mossberger 2018; van Deursen and van Dijk 2009, 2010, 2015; van Dijk and Hacker 2003). The possession of adequate digital skills has been found to have a positive effect on internet usage (van Deursen, Helsper, and Eynon 2014; van Deursen and van Dijk 2015), and particularly on expressive uses (Shaw and Hargittai 2018).

*Internet usage* (IU). Internet usage has been defined in various ways, including frequency of use, connection duration, diversity of benefits, and variety of activities (Hargittai, and Hinnant 2008; van Deursen and van Dijk 2014). Internet usage has been categorized in terms of information seeking, personal development, social interaction, leisure and entertainment, commercial transactions, trade, emailing, blogging, and production (Blank and Groselj 2014; van Deursen and van Dijk 2014). Perhaps a more important outcome than the amount of time spent online is precisely how time online is used, with such outcomes classified (Helsper, van Deursen, and Eynon 2016) as follows: economic (related to property, finance, employment, education, etc), cultural (related to identity, belonging, cultural expressions, etc), social (reflecting personal, formal, political networks, etc), and personal (health, lifestyle, self-realization, leisure, etc.).

Our model, an adaption of that proposed by van Dijk and van Deursen (2015) and Lamberti et al. (2021), is depicted in Figure 1. It focuses specifically on PA, DS, and IU as the more immediately measurable constructs (i.e., our use of secondary data does not allow us to consider ATT as an antecedent of PA).

[Insert Figure 1. The sequential model of internet use about here.]

### 2.1.3. The sequential model: cumulative and recursive

So far, we have described the constructs underpinning the sequential model of internet appropriation that produce different digital divide levels. However, even though the causality model that explains the third digital divide (arising from social and cultural advantages/disadvantages) is sequential, the RA theory also indicates that the model is, additionally, cumulative and recursive (van Dijk 2002, 2006), and apply equally to older and newer media.

The sequential levels of the digital divide are cumulative in two ways. First, cumulative waves of innovation have led in turn to desktop computers, the internet, laptops, netbooks, tablets, smartphones, and many other spin-off devices. Devices are further enhanced by network innovations, e.g., smartphone usefulness has been enhanced by the advent of 3G, 4G and 5G data technologies. Devices differ in terms of cost, typically related to ease of use, operating systems, and the innovativeness of embedded technologies: more costly devices are faster, easier to use, and have better software, which, ultimately, affects what the individual can do and the derived benefits. Second (and related to this last point), individuals adopt the new technologies at different temporal points (Norris 2001; van Dijk 2002, 2006) and at different speeds. If the pattern of adoption of digital technologies were to follow Norris' normalization hypothesis (Norris 2001) suggesting that social differences could eventually fade away, all individuals would eventually achieve the same level of adoption. However, evidence for Europe, as reported by Norris (2001), supports a social stratification thesis, i.e., the persistence of social differences that are reflected in the digital divide. Thus, society is stratified into privileged and underprivileged social groups that achieve a different level of adoption at different times in line with their resources, and the outcome is an increase in divergence in life chances, access to capital and resources, and to privileged social positions.

The sequential model is also recursive. With each new digital innovation, the adoption process starts anew and runs concurrently with previous adoption processes, although the skills, use, and benefits may be fundamentally grounded in previous experiences. The fact that many new innovations build on a previous innovation makes it easier for people who have already adopted a technology to foresee the value of an enhancement. Others left behind will see their digital skills stagnate (i.e., the skill difference will be greater in both absolute and relative terms), and they will not enjoy the social and economic benefits of the more advanced technology. The skill deficit, in turn, may reduce their motivation to adopt the new digital innovation. Furthermore, each advance in digital devices typically combines with other complementary innovations to reinforce productivity, e.g., smartphones with advanced printers, television sets with internet connections, laptops with long-lasting batteries and smartphones sharing access to computer networks, both local and global, and shared broadband access at home or in the office and the new WiFi mesh technologies.

The cumulative and recursive properties of the model are shown in Figure 2 (red lines), which depicts what we call the cyclic model of internet appropriation. The cumulative and recursive properties would suggest that the sequential effect of the different digital divides is underestimated when we use cross-sectional data because cyclic effects are not reflected in the sequential model.

[Insert Figure 2. The cyclic model of internet appropriation about here.]

## 3. Method

### 3.1. PLS-SEM

PLS-SEM (Hair et al. 2016; Wold 1985) connects a set of observed variables (i.e., indicators or manifest variables) with latent variables (i.e., constructs) through a system of linear relationships (Hair et al. 2016) that are simultaneously quantified by applying a set of sequential multiple linear regressions. Figure 3 depicts the model specifications, involving computation of two models: an outer model that relates manifest variables (MVs) to latent variables (LVs) in a reflective manner, and an inner model that reflects the strength and direction of relationships between the LVs in a reflective or formative manner. The relationship is reflective when it is hypothesized that an LV causes the observed association with an indicator, while it is formative when an LV is generated by the indicator (in the form of a scale). Reflective relationships, but not formative relationships, need to be highly correlated as each indicator describes a different aspect of the LV.

[Insert Figure 3. PLS-SEM model specifications about here.]

Referring to Figure 3, let us assume that we have $P$ indicators that are observed on $N$ units $(i = 1, ..., n, ..., N)$, then the data $x_{npk}$ are collected in a partitioned matrix $X = [X_1, ..., X_k, ..., X_K]$ where $x_{pk}$, with $p = 1, ..., p_k$, and $\sum_{k=1}^{K} p_k = P$ is an indicator belonging to the $k$-th block $X_k$. We also assume that a LV $\xi_k$ is associated to each block $X_k$.

The reflective blocks $X_1$ and $X_3$ and the formative block $X_2$ in the outer model are formalized in Eqs. (1) and (2), respectively:

$$x_{pk} = \lambda_{pk0} + \lambda_{pk}\xi_k + \varepsilon_{pk}, \qquad (1)$$

$$\xi_k = \pi_{pk0} + \sum_k \pi_{pk} x_{pk} + \delta_{pk}, \qquad (2)$$

where, $\lambda_{pk0}$ and $\pi_{pk0}$ are both location parameters, $\lambda_{pk}$ is the loading coefficient that captures the effect of $\xi_k$ on $x_{pk}$, while, for the formative case, $\pi_{pk}$ represents the weight of the linear combination associated with the indicator $x_{pk}$, and $\varepsilon_{pk}$ *and* $\delta_{pk}$ are measurement error variables.

Finally, a generic dependent LV is linked to the corresponding explanatory LVs by the following equation:

$$\xi_k = \beta_{k0} + \beta_{kk'} \xi_{k'} + \zeta_k, \qquad (3)$$

where $\beta_{kk'}$ is what is known as a path coefficient, which captures the effects of the predictor LV $\xi_{k'}$ on the dependent LV $\xi_k$, and where $\zeta_k$ is the inner residual variable.

Concerning validation, several tests can assess the quality of the outer model. Reliability is most frequently tested using Cronbach's α, composite reliability (CR), Dijkstra-Henseler's $\rho$, and average variance expected (AVE). As for the inner model, this is validated by analysing the length and significance of path coefficients and using $R^2$. Further details and a comprehensive list of possible tests are available in Hair et al. (2019), Hair et al. (2016), and Espostio Vinzi et al. (2010).

**3.2. Cyclic effects in PLS-SEM**

Consider a simple model with two LVs, one exogenous (i.e., predictive ($\xi_1$)) and the other endogenous (i.e., dependent ($\xi_2$)), reflecting antecedents and consequences. Note that, for the sake of simplicity, we do not differentiate between exogenous and endogenous in the notation. Under the classical causal model paradigm, we have a causal relationship between $\xi_1$ and $\xi_2$ when $\xi_1$ affects $\xi_2$ ($\xi_1 \to \xi_2$); this effect is sequential, as we can establish a temporal order between the two LVs, such that $\xi_1$ is antecedent to $\xi_2$ (Figure 4, left). Another situation is when $\xi_1$ affects $\xi_2$ and when $\xi_2$ also affects $\xi_1$ ($\xi_1 \rightleftarrows \xi_2$), i.e., $\xi_1$ and $\xi_2$ are involved in a feedback loop; in this case, the influence is cyclic (Figure 4, right).

[Insert Figure 4. Simple causality path model about here.]

Considering the model of internet appropriation (van Dijk 2005, 2015, 2020) and the relationship between the constructs, we can intuitively infer a sequential effect: if people have opportunities to physically access the internet (PA), they will develop better digital skills (DS), and this will increase usage (IU). However, just as intuitively, we can also infer a cyclic effect:

more use of the internet (IU) will result in more physical access (PA) and better digital skills (DS).

In PLS-SEM (as mentioned earlier), we face the problem of estimating the cyclic effect when working with cross-sectional data. In this situation, we do not have the same indicators measured for the same individuals at different time points. Indeed, we can say that measurement of the variables at another time point is latent. However, let us assume that cyclic effects exist (as intuited above in relation to internet appropriation and its constructs) and that we want map the entire relationship structure. In this case, we account for any cyclic effect by using available variables to generate a proxy of variables measured at different time points.

### 3.3. The proposal: two-step approach

Our proposal for estimating cyclic effects for cross-sectional data involves two steps (Figure 5). First, to identify sequentiality in the LVs, we estimate the model applying the classical PLS-SEM procedure, estimating the coefficients and dependent LV scores. Second, we estimate cyclic effects in a separate model where the dependent LV is now used to predict antecedent variables using, instead of the original LV indicators, the dependent LV scores.

[Insert Figure 5. Two-step approach to estimating cyclic effects about here.]

Application to internet appropriation and the PA, DS, and IU constructs (Figure 6) is as follows: in the first model, sequentiality is established for PA, DS, and IU and scores are calculated for the coefficients that sequentially relate the three constructs; and in the second model, IU is now established as an antecedent of both PA and DS, and the IU score obtained in the first model now yields new coefficients that quantify the cyclic effects.

[Insert Figure 6. Two-step approach to estimating cyclic effects in internet appropriation about here.]

### 3.3.1. Cyclicity reinforcement testing

Determining cyclic causality requires not only quantifying effects (length and significance) but also determining the relative significance of cyclic effects versus sequential effects: if the cyclic effect is significantly higher respect the sequential, we could state that there is a reinforcement effect. Considering the model of internet appropriation (van Dijk 2005, 2015, 2020), if the cyclic effect of IU on PA, and DS is significant higher, we can assume that the sequential effect of PA and DS on IU reinforces the effect of IU on PA and DS.

For this purpose, we adapted the parametric test proposed by Keil et al. (2000) to test the significance of a categorical variable for a path coefficient estimated using PLS-SEM[1].

Following the Keil et al. (2000), and considering a simple model with two LVs (Figure 4, right), we can compare significant differences between sequential and cyclic effects (SE and CE in the equations below) by comparing differences between two coefficients and using the bootstrapping procedure to estimate standard error. Adapting the parametric test, the null hypothesis is that there is no difference between effects (i.e., $H_0: \beta_{SE}=\beta_{CE}$), while the alternative hypothesis is that CE effect is significantly higher than the SE effect (i.e., $H_1: \beta_{SE} > \beta_{CE}$). The null hypothesis is tested, and *df* is determined, as follows:

$$t = \frac{|\beta_{SF} - \beta_{CF}|}{\sqrt{\frac{(n-1)}{n}\left(\sigma^2_{\beta_{SF}} + \sigma^2_{\beta_{CE}}\right)}} \quad . \tag{4}$$

$$df = \left\| \sqrt{\frac{\left(\frac{(n-1)}{n}(\sigma^2_{\beta_{SE}} + \sigma^2_{\beta_{CF}})\right)^2}{\frac{(n-1)}{n^2}(\sigma^4_{\beta_{SE}} + \sigma^4_{\beta_{CF}})}} \right\| \quad . \tag{5}$$

## 4. Illustration: Accounting for the cyclic effect

In this section we show how to estimate the cyclic effect by applying the two-step approach to the sequential model of internet appropriation. Below we briefly present our data, and then describe the results of the analysis. Data, measurements and model are a based on the analysis proposed by Lamberti et al. (2021).

### 4.1. Data and measurements

Representative data on ICT usage in 2016 were taken from a Eurostat EU27+UK survey[2]. Selected were 151 660 individuals aged over 15 years who had used the internet in the previous three months. Figure 7 reports gender, age, education, and employment percentages for the sample (just over half were female, three quarters were aged 25–64 years, a third had a high education level, and two thirds were employed).

[Insert Figure 7. Sample characteristics about here.]

Measurement of the RA theory indicators (PA, DS, and IU) is briefly summarized below (full details are available in Appendix A1).

PA was measured in terms of a yes/no response regarding the use of six device types (desktop computer, laptop/notebook, tablet, mobile phone/smartphone, smart TV, and other mobile devices, e.g., e-book reader, smartwatch). Due to the dichotomous nature of the indicators, PA was operationalized through multiple correspondence analysis (MCA) (Greenacre and Blasius, 2006), with the coordinates for the first dimension reflecting a PA intensity scale. Figure 8, which depicts the two first MCA dimensions, shows variables coloured by contribution (darker shades reflect greater contributions). The first dimension, reflecting PA intensity and accounting for 76% of the total variance, contrasts non-use and use of particular devices (left- and right-hand side of the plot, respectively), and showing that smartphones and tablets make the greatest contribution to this dimension.

[Insert Figure 8. MCA factorial map of physical access intensity about here.]

DS was measured in terms of four abilities – to use communication applications, networks and digital devices to access and manage information; to use technology to communicate with others; to use technology for problem-solving tasks; and use and knowledge of software – labelled as information skills, communication skills, problem-solving skills, and software skills, respectively. Self-reported skill levels, measured on a four-point Likert scale, ranged from 1 (no skills) to 4 (advanced skills).

Finally, IU was measured in terms of online activities (including emailing, reading news, playing games, listening to music, managing a website, running a business, etc), grouped into four categories (as adapted from van Deursen and van Dijk 2014, and Blank and Groselj 2014) labelled social interaction, information-seeking, leisure, and commercial transactions. Self-reported activity levels, measured on a four-point Likert scale, ranged from 1 (not used) to 4 (highly frequent use).

## 4.2. Results
### 4.2.1. Sequential model results

PA was a single-item scale (as the indicators were dichotomous, the first MCA dimension was used to build the scale), so no additional validation was needed when the measurement model was analysed (MCA scales are optimal scales according to Greenacre and Blasius 2006)). DS and IU were modelled assuming that each was an antecedent of its own indicators (i.e., we adopted a reflective approach) and were validated by checking CR indexes (Esposito Vinzi et al. 2010; Hair et al. 2016). Results, reported in Appendix A2, are consistent with evidence reported by Lamberti et al. (2021).

Evidence supporting sequential causality is depicted in Figure 9, which shows path coefficient and $R^2$ results for the inner model (significance according to a 95% confidence interval (CI) is indicated by asterisks). Our results indicate that PA strongly influences DS development (β=0.458) and more weakly influences IU (β=0.240), while the most significant influence on IU is DS (β=0.647).

[Insert Figure 9. Sequential model estimation about here.]

### 4.2.2. Cyclic model results

Figure 9 only depicts a partial image of reality, showing that PA has effects on DS and IU, and DS has an effect on IU. However, as depicted in Figure 10 (95% CI significance indicated with asterisks), the influence of the constructs is also cyclic (lines and coefficients in red), as IU also significantly affects, in a reciprocal manner, both PA (β=0.537) and DS (β=0.761).

[Insert Figure 10. Cyclic model estimation about here.]

Table 1 reports the parametric test results for the comparative significance of the cyclic effect versus the sequential effect for PA, DS, and IU, showing the impact of the path coefficients, differences in absolute value, and the corresponding t-statistics and p-values, and confirming that there is a reinforcement effect: the cyclic coefficients (IU on PA, and IU on DS) are significantly higher than the sequential coefficients (PA on IU, and DS on IU).

[Insert Table 1. Parametric test results about here.]

## 5. Discussion

We have proposed a new, easily implemented, and easily interpreted two-step approach to dealing with cyclic effects in a causal model when working with cross-sectional data using a PLS algorithm. It is based on, first, estimating the classical causal model, and second, using the resulting dependent variable scores to estimate the cyclic effects. The novelty of the method is that it can be applied to cross-sectional datasets consisting of data measured at a specific moment in time.

The sequential model of internet appropriation proposed by the RA theory (van Deursen and van Dijk 2015; van Dijk 2005, 2020) to explain the digital divide only paints a partial picture of the social mechanisms that underpin internet appropriation, based as it is on an acyclic hypothesis. To demonstrate the importance of considering cyclic effects, using secondary ICT data from Eurostat we built the sequential model (van Deursen and van Dijk 2015; van Dijk 2005, 2020), measured the PA, DS, and IU constructs, and analysed the sequential causality between those constructs. This enabled us to determine how and to what extent PA and DS affect IU. In our analysis of effects, we found, as acyclic effects that corroborate previous research (Hargittai 2002; Lamberti et al. 2021; van Deursen and Helsper 2015), that PA is vital for DS but less vital for IU, and that IU is strongly determined by DS, and, as cyclic effects, that IU enhances both PA and DS.

## 5.1. Methodological contribution

Our approach represents a significant methodological contribution as it is the first attempt to tackle cyclicity in causal models using PLS-SEM and cross-sectional data measured at a specific moment in time. Our two-step approach represents an advance in the analysis of causal relationships between LVs, as it enables study of the entire structure of causality, i.e., both acyclic and cyclic. This contribution is especially relevant, given that, for the same group of observations, longitudinal datasets are generally less available than cross-sectional datasets.

## 5.2. Sociological contribution

From a sociological point of view, our proposal can enhance studies of internet appropriation, as, compared to the sequential model (van Deursen and van Dijk 2015; Lamberti et al. 2021, among the most recent), the cyclic model provides a more complete view of the interplay among the social mechanisms that underpin internet appropriation. Specifically, we found, in support of Norris' social stratification thesis (2001), that not only is PA crucial for DS and that both have an effect on IU, but also that IU has a positive impact on both PA and DS.

## 5.3. Limitations

Our approach is not without limitations. First, the estimated coefficients for the cyclic effects are merely a proxy of the real coefficients, given that we use the same set of variables for both the acyclic and cyclic effects. This limitation originates in the fact that we use cross-sectional data and so do not have same variables measured at different moments in time. Second, using the same set of variables to estimate acyclic and cyclic effects produces a bias in coefficient estimation, as cyclic effects depend on the indirect effect of the predictor variables on the dependent variable (in our case, PA and DS on IU). Thus, since we use PA and DS to estimate IU and then use the IU score to estimate the return effect on PA and DS, we can expect the cyclic effect to be higher than it really is. However, this bias can theoretically be justified if we assume, based on intuition, that variables influence each other in a reciprocal manner. Third, this method can only be applied when there is at least one intermediate LV in the model (in our case DS), i.e., cyclic effects cannot be estimated for just two constructs. This limitation is due to how PLS-SEM estimates parameters. Since PLS-SEM maximizes the variance of a block of indicators related to each LV, a causal relationship is not considered when there are just two LVs; to illustrate, if we only considered DS and IU as variables in the sequential model, we would obtain the same correlation coefficient for estimates of both acyclic and cyclic effects. A final limitation is related to our use of secondary data, as it meant that we could not consider ATT as an antecedent of PA, as posited by the RA theory, and that scale measurement may have been affected by the fact that the indicators available were not as rich as the indicators obtained from surveys designed specifically to test a particular theory, as done by van Deursen and van Dijk (2015).

## 6. Conclusion

Working with SEM and cross-sectional data, and depending on the studied phenomenon, assuming an acyclic model may mean that we obtain only a partial view of the mechanisms that explain causal relationships between a set of theoretical constructs, given that variables are treated as antecedents and consequences. Our two-step approach allows researchers to identify and measure cyclic effects when working with cross-sectional data and a PLS modelling algorithm. We demonstrate the importance of considering cyclic effects using RA theory and the sequential model of internet appropriation. Our results show that opportunities for physical access followed by digital skills acquisition enhance internet usage (acyclic effects), but also that internet usage intensity, in reverse, reinforces both digital skills and physical access (cyclic effects), supporting Norris' (2001) social stratification hypothesis regarding future evolution of the digital divide.

**References**

Asparouhov, Tihomir, Ellen L. Hamaker, and Bengt Muthén. 2018. "Dynamic Structural Equation Models." *Structural Equation Modeling: A Multidisciplinary Journal* 25(3): 359-388.

Attewell, Paul A. 2001. "Social Inclusion and Equity in Modern Information and Knowledge Societies." *Sociology of Education* 74:252-259.

Becker, Jan M., Kristina Klein, and Martin Wetzels. 2012. "Hierarchical Latent Variable Models in PLS-SEM: Guidelines for Using Reflective-formative Type Models." *Long Range Planning* 45(5-6): 359-394.

Broos, Agnetha, and Keith Roe. 2006. "The Digital Divide in the Playstation Generation: Self-efficacy, Locus of Control and ICT Adoption among Adolescents." *Poetics* 34(4–5): 306-317. DOI:

Blank, Grant, and Darja Groselj. 2014. "Dimensions of Internet Use: Amount, Variety, and Types." *Information Communication, and Society* 17(4): 417-435.

Crocetta, Corrado, Laura Antonucci, Rosanna Cataldo, Roberto Galasso, Maria G. Grassia, Carlo N. Lauro, and Marina Marino. 2021. "Higher-order PLS-PM Approach for Different types of Constructs." *Social Indicators Research* 154(2): 725-754.

DiMaggio, Paul, Eszter Hargittai, Coral Celeste, and Steven Shafer. 2004. "From Unequal Access to Differentiated Use: A Literature Review and Agenda for Research on Digital Inequality." Pp. 355-400 in Social Inequality, New York: Russell Sage Foundation.

Drton, Mathias, Christopher Fox, and Samuel Y. Wang. 2019. "Computation of Maximum Likelihood Estimates in Cyclic Structural Equation Models." *The Annals of Statistics* 47(2): 663-690.

Esposito Vinzi, Vincenzo, Wynne W. Chin, Jörg Henseler, and Huiwen Wang. 2010. *Hand Book of Partial Least Squares: Concepts, Methods and Applications*. Berlin Heidelberg: Springer

Ferrari, Anusca 2012. "*Digital Competence in Practice: An Analysis of Frameworks*." Technical Report by the Joint Research Centre of the European Commission. Publications Office of the European Union. Retrieved June 1, 2022 (https://op.europa.eu/en/publication-detail/-/publication/2547ebf4-bd21-46e8-88e9-f53c1b3b927f/language-en)

Greenacre, Micheal, and Jorg Blasius. 2006. *Multiple Correspondence Analysis and Related Methods*. New York: Chapman and Hall/CRC.

Hair, Joseph F., Jeffrey J. Risher, Marko Sarstedt, Christian M. Ringle. 2019. "When to Use and How to Report the Results of PLS-SEM." *European Business Review* 31(1): 2-24.

Hair, Joseph F., Marko Sarstedt, Christian M. Ringle, Siegfried P. Gudergan. 2017. *Advanced Issues in Partial Least Squares Structural Equation Modeling*. Los Angeles: Sage.

Hair, Joseph F., Tomas M. Hult, Christian M. Ringle, and Marko Sarstedt. 2016. *A Primer on Partial Least Squares Structural Equation Modeling (PLS-SEM)*. Sage.

Hargittai, Eszter, and Amanda Hinnant. 2008. "Digital Inequality: Differences in Young Adults' Use of the Internet." *Communication Research* 35(5): 602–621.

Hargittai, Eszter, Anne M. Piper, and Meredith R. Morris. 2018. "From Internet Access to Internet Skills: Digital Inequality among Older Adults." *Universal Access in the Information Society* 18: 881-890.

Hargittai, Eszter, and Steven Shafer. 2006. "Differences in Actual and Perceived Online Skills: The Role of Gender." *Social Science Quarterly* 87(2): 432-448.

Hargittai, Eszter. 2003. "The Digital Divide and What to Do About It." Pp: 822-841 in *New Economy Handbook* edited by D. C. Jones, San Diego, CA: Academic Press.

Hargittai, Eszter. 2001. "Second-level Digital Divide: Mapping Differences in People's Online Skills." Retrieved June 1, 2022 (arXiv preprint cs/0109068).

Helsper, Ellen. J., Alexander J.A.M. van Deursen, and Rebecca Eynon. 2016. Measuring Types of Internet Use from: Digital Skills to Tangible Outcomes Project Report. Retrieved June 1, 2022 (https://www.alexandervandeursen.nl/Joomla/Media/Reports/2016%20-%20Report_Internet_uses.pdf)

Hyttinen, Antti, Frederick Eberhardt, and Patrik O. Hoyer. 2012. "Learning Linear Cyclic Causal Models with Latent Variables." *The Journal of Machine Learning Research* 13(1): 3387-3439.

Keil, Mark, Bernard C. Y. Tan, Kwok-Kee Wei, Timo Saarinen, Virpi Tuunainen, and Arjen Wassenaar. 2000. "A Cross-cultural Study on Escalation of Commitment Behavior in Software Projects." *MIS quarterly* 24(2): 299-325.

Kuhlemeier, Hans, and Bas Hemker. 2007. "The Impact of Computer Use at Home on Students' Internet Skills." *Computers, and Education* 49(2): 460–480.

Lamberti, Giuseppe, Jordi Lopez-Sintas, and Jakkapong Sukphan. (2021). "The Social Process of Internet Appropriation: Living in a Digitally Advanced Country Benefits Less Well-educated Europeans." *Telecommunications Policy* 45(1): 102055.

Litt, Eden. 2013. "Measuring Users' Internet Skills: A Review of Past Assessments and a Look Toward the Future." *New Media & Society* 15(4): 612–630.

Livingstone, Sonia, and Ellen J. Helsper. 2007. "Gradations in Digital Inclusion: Children, Young People and the Digital Divide." *New Media & Society* 9(4): 671–696.

Mossberger, Karen, Caroline J. Tolbert, Allison Hamilton. 2012. "Measuring Digital Citizenship : Mobile Access and Broadband." *International Journal of Communication* 6: 2492–2528. Retrieved June 1, 2022 (https://ijoc.org/index.php/ijoc/article/view/1777)

Napoli, Philip M., and Jonathan A. Obar. 2014. "The Emerging Mobile internet Underclass: A Critique of Mobile Internet Access." *Information Society* 30(5): 323–334.

Norris, Pippa. 2001. *Digital Divide: Civic Engagement, Information Poverty, and the Internet Worldwide*. Cambridge: Cambridge University Press.

Parsons, Cheryl, and Steven F. Hick. 2008. "Moving from the Digital Divide to Digital Inclusion." *Currents: Scholarship in the Human Services 7*(2): Article 2 Retrieved June 1, 2022 (https://journalhosting.ucalgary.ca/index.php/currents/article/view/15892)

Shaw, Aaron, and Ezter Hargittai. 2018. "The Pipeline of Online Participation Inequalities: The Case of Wikipedia Editing." *Journal of Communication* 68(1): 143–168.

Sarstedt, Marko, Joseph F. Hair, Cheah Jun-Hwa, Jan-Michael Becker, and Christian M. Ringle. 2019. "How to Specify, Estimate, and Validate Higher-Order Constructs in PLS-SEM." *Australasian Marketing Journal* 27(3): 197-211.

Sarstedt, Marko, Jörg Henseler, and Christian M. Ringle. 2011. "Multigroup Analysis in Partial Least Squares (PLS) Path Modeling: Alternative Methods and Empirical Results." Pp. 195-218 in Measurement and Research Methods in International Marketing edited by Sarstedt, M., Manfred Schwaiger, and Charles R. Taylor. Bingley: Emerald Group Publishing Limited.

Robinson, Laura. 2009. "A Taste for the Necessary: A Bourdieuian Approach to Digital Inequality." *Information, Communication, and Society* 12(4): 488-507.

Ullman, Jodie B., and Peter M. Bentler. 2012. *Structural Equation Modeling*. Handbook of Psychology, Second Edition, 2.

van Deursen, Alexander J. A. M., Cedric Courtois, and Jan A. G. M. van Dijk. 2014. "Internet Skills, Sources of Support, and Benefiting from Internet Use." *International Journal of Human-Computer Interaction* 30(4): 278–290.

van Deursen, Alexander J. A. M.., and Ellen J. Helsper. 2015. "The Third-level Digital Divide: Who Benefits Most from Being Online?" Pp. 29-52 in Communication and Information Technologies Annual (Studies in Media and Communications, Vol. 10). Bingley: Emerald Group Publishing Limited.

van Deursen, Alexander J. A. M., Ellen J. Helsper, and Rebecca Eynon. 2016. "Development and Validation of the Internet Skills Scale (ISS)." *Information Communication, and Society* 19(6): 804–823.

van Deursen, Alexander J. A. M., Ellen J. Helsper, and Rebecca Eynon. 2014. "Measuring Digital Skills: From Digital Skills to Tangible Outcomes Project Report." Retrieved October 07, 2014 from (http://www.oii.ox.ac.uk/research/projects/?id=112)

van Deursen, Alexander J. A. M., and Karen Mossberger. 2018. "Any Thing for Anyone? A New Digital Divide in Internet-of-Things Skills." *Policy and Internet* 10(2): 122–140.

van Deursen, Alexander J. A. M., and Jan A. G. M. van Dijk. 2019. "The first-level Digital Divide Shifts from Inequalities in Physical Access to Inequalities in Material Access." *New Media & Society* 21(2): 354–375.

van Deursen, Alexander J. A. M., and Jan A. G. M. van Dijk. 2015. "Toward a Multifaceted Model of Internet Access for Understanding Digital Divides: An Empirical Investigation." *The Information Society* 31(5): 379–391.

van Deursen, Alexander J. A. M., and Jan A. G. M. van Dijk. 2014. "The Digital Divide Shifts to Differences in Usage." *New Media & Society* 16(3): 507–526.

van Deursen, Alexander J. A. M., and Jan A. G. M. van Dijk. 2010. "Internet Skills and the Digital Divide." *New Media & Society* 13(6): 893–911.

van Deursen, Alexander J. A. M., and Jan A. G. M. van Dijk. 2009. "Using the Internet: Skill Related Problems in Users' Online Behaviour." *Interacting with Computers* 21(5–6): 393–402.

van Deursen, Alexander J. A. M., and Jan A. G. M. van Dijk. 2008. Measuring digital skills: Performance tests of operational, formal, information and strategic Internet skills among the Dutch population. In 58th Conference of the International Communication Association, Montreal, Canada May 22-26, (pp. 1–25)

van Deursen, Alexander J. A. M., Jan A. G. M. van Dijk, and Oscar Peters. 2011. "Rethinking Internet Skills: The Contribution of Gender, Age, Education, Internet Experience, and Hours Online to Medium- and Content-related Internet Skills." *Poetics* 39(2): 125–144.

van Dijk, Jan A. G. M. 2020. *The Digital Divide*. Cambridge, UK ; Malden, MA: Polity Press.

Van Dijk, Jan A. G. M. 2012. "The Evolution of the Digital Divide-the Digital Divide Turns to Inequality of Skills and Usage." Pp. 57-75 in *Digital Enlightenment Yearbook 2012* edited by Bus, Jacques; Mireille Hildebrandt, and Malcolm Crompton IOS Press.

van Dijk, Jan A. G. M. 2005. *The Deepening Divide: Inequality in the Information Society*. SAGE publications.

van Dijk, Jan A. G. M. 2006. "Digital Divide Research, Achievements and Shortcomings." *Poetics 34*(4–5): 221–235.

van Dijk, Jan A. G. M. 2002. "A Framework for Digital Divide." *The Electronic Journal of Communication*, *12*(1-2):7.

van Dijk, Jan A. G. M., and Kennet Hacker 2003. "The Digital Divide as a Complex and Dynamic Phenomenon." *The Information Society: An International Journal* 19(4): 315–326.

Venkatesh, Viswanath, Michael G. Morris, Gordon B. Davis, and Fred D. Davis 2003. "User Acceptance of Information Technology: Toward a Unified View." *MIS Quarterly* 27(3): 425.

Wold, Herman. 1985. "Partial least squares". In Encyclopedia of Statistical Sciences edited by Kotz, Samuel, Campbell B. Read, N. Balakrishnan, Brani Vidakovic, and Norman L. Johnson. John Wiley, and Sons.

Zillien, Nicole, and Ezter Hargittai. 2009. "Digital Distinction: Status-Specific Types of Internet Usage." *Social Science Quarterly* 90(2): 274-291.

**Footnotes**

[1]The parametric test is based on a bootstrap re-sampling procedure to evaluate coefficient differences. The bootstrap procedure is used to estimate the standard errors of the path coefficients. Finally, the difference between coefficients is tested using t-statistic (for details see Hair et al., 2017).

[2] Details of data collection procedure are available at: https://circabc.europa.eu/ui/group/4f80b004-7f0a-4e5a-ba91-a7bb40cc0304/library/8bc71641-bd53-4039-b9f0-71d87822749d/details

**Appendix A1. Constructs definition and operationalization**

For each construct, we provide the definition, the indicators used to measure it, how it was operationalized, references, and a comparison with the construct employed in the van Deursen and van Dijk (2015) model.

[Insert Table A1. Constructs definition and operationalization about here.]

**Appendix A2. Outer model validation**

DS and IU were validated by checking four common reliability indexes (Esposito Vinzi et al., 2010; Hair et al., 2019): to measure internal consistency (which should be greater than 0.7), Cronbach's $\alpha$, Dillon's $\rho$, and Dikstra's $\rho$, and to measure unidimensionality, the difference between the first and second eigenvalues, with only the first eigenvalue expected to be greater than 1. We also checked loading significance according to bootstrap intervals (calculated with 500 repetitions) and their length, which, in the case of reflective indicators, should be greater than 0.7, and the average variance extracted (AVE), which should be greater than 0.5 (indicating that the constructs reflect at least 50% of the variance in the indicators). Results are reported in Table A1. Cronbach's $\alpha$ and Dijkstra's $\rho_A$ were above the threshold of 0.7 for DS and almost reached the threshold for IU, and Dillon's $\rho$ was high for both constructs. Furthermore, for both DS and IU, only the first eigenvalue of the PCA was above 1, indicating evidence favouring construct unidimensionality. All item loadings were close to or greater than the 0.7 threshold and were significant according to the 95% CI, while AVE values were greater than 0.5 for both DS and IU.

[Insert Table A2. Scale validation about here.]

**Table**

Table 1. Parametric test results.

| Effects | SE | CE | Abs (diff.) | t-statistic | p-value |
|---|---|---|---|---|---|
| PA⇌IU | 0.24 | 0.537 | 0.297 | 126.81 | 0.002 |
| DS⇌IU | 0.647 | 0.761 | 0.114 | 75.58 | 0.004 |

SE, sequential effect; CE, cyclic effect.

Table A1. Constructs definition and operationalization.

| Construct | Definition | Items | Operationalization | Reference | van Deursen and van Dijk' model |
|---|---|---|---|---|---|
| PA | Opportunity to access the internet | *Which of the following devices have you used to access internet?* <br><br> (1) Desktop computer, (2) Laptop or netbook, (3) Tablet computer, (4) Mobile phone or smartphone, (5) Other mobile device (e.g., e-reader, smartwatch), (6) Smart TV (directly connected to the internet) | First dimension of multiple correspondence analysis (MCA) | van Deursen and van Dijk 2015 | Material Internet Access (single-item scale), measured dichotomously using 7 questions regarding devices used to access the internet: desktop PC, laptop PC, tablet PC, smartphone, game console, TV, e-reader. |
| DS | Ability to use the internet | *Please indicate level of the following skills:* <br><br> (1) Obtain information, (2) Communicate information, (3) Solve software and hardware problems, (4) Solve substantive problems | Eurostat Likert scale, ranging from 1 (no skills) to 4 (highest skills) | Ferrari 2012; Hargittai et al. 2018; van Deursen and Mossberger 2018; van Deursen and van Dijk, 2009, 2010, 2015; van Dijk and Hacker 2003 | Medium- and Content-related Internet Skills (single-item scale) |
| IU | Number and variety of different internet activities participated in online | *For which of the following activities have you used internet?* <br> (1) **Social interaction**: 1 – Sending/receiving e-mails; 2 - Telephoning over the internet/video calls (via webcam) over the internet (e.g., Skype or Facetime); 3 - Participating in social networks (creating user profiles, posting messages or other contributions to Facebook, etc); 4 - Uploading self-created content (text, photos, music, videos, software, etc) to any website for sharing <br> (2) **Information-seeking**: 5 - Reading online news/newspapers/news magazines; 6 - Finding information about goods or services; 7 - Seeking health-related information (e.g., injury, disease, nutrition, health, etc); 8 - Making an appointment with a practitioner via the website (e.g., hospital or healthcare centre) <br> (3) **Leisure**: 9 - Listening to music (e.g., web radio, music streaming) 10 - Watching internet-streamed TV (live or catch-up) from broadcasters; 11 - Watching video on demand from commercial services (Netflix, HBO, etc); 12 - Playing video games <br> (4) **Commercial transactions**: 13 - Using services related to travel or travel-related accommodation; 14 - Selling goods or services, e.g., via auctions (e.g. eBay); 15 - Internet banking; 16 - Using payment accounts (e.g. PayPal) to pay for goods or services purchased online | 16 dichotomous items reflecting a broad range of internet activities, grouped and summed in 4 categories: social interaction, information-seeking, leisure, commercial transaction | Blank and Groselj 2014; van Deursen and van Dijk 2014; van Deursen and van Dijk 2015 | Internet Usage (single-item scale), measuring frequency of engagement in 21 activities, with items summed into a single scale that reflected diversity of usage activities (0 to 21) |

Table A2. Scale validation

| LV | Indicator | Cronbach's $\alpha$ | CR | PCA eigen. 1 | PCA eigen. 2 | Dijkstra's $\rho_A$ | AVE | Orig. | 95% CI | |
|---|---|---|---|---|---|---|---|---|---|---|
| PA | Physical access (MCA dim. 1) | | | | | | | 1 | | |
| DS | | 0.752 | 0.844 | 2.310 | 0.700 | 0.758 | 0.575 | | | |
| | Information | | | | | | | 0.725 | 0.721 | 0.728 |
| | Communication | | | | | | | 0.695 | 0.692 | 0.697 |
| | Problem-solving | | | | | | | 0.839 | 0.837 | 0.841 |
| | Software | | | | | | | 0.768 | 0.765 | 0.771 |
| IU | | 0.694 | 0.813 | 2.090 | 0.797 | 0.696 | 0.521 | | | |
| | Social | | | | | | | 0.732 | 0.729 | 0.734 |
| | Information | | | | | | | 0.693 | 0.689 | 0.696 |
| | Leisure | | | | | | | 0.725 | 0.723 | 0.728 |
| | Comm. trans. | | | | | | | 0.736 | 0.733 | 0.738 |

Figure 1. The sequential model of internet use

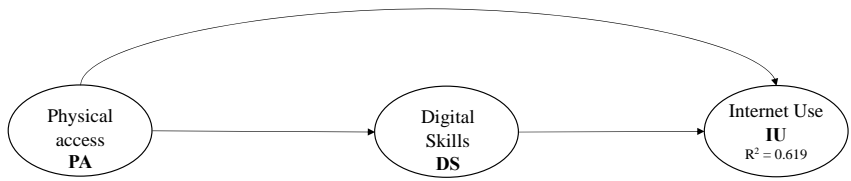

Figure 2. The cyclic model of internet appropriation.

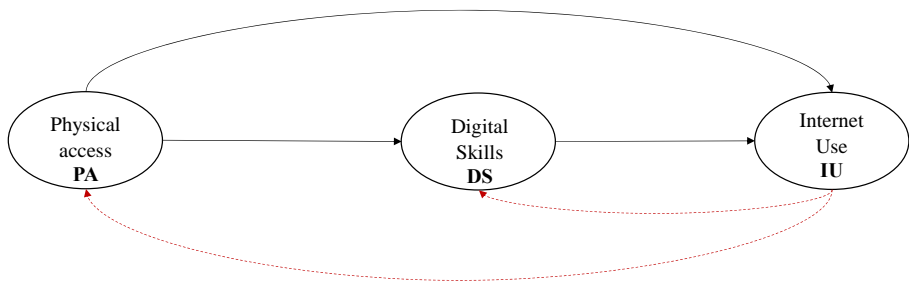

Figure 3. PLS-SEM model specifications.

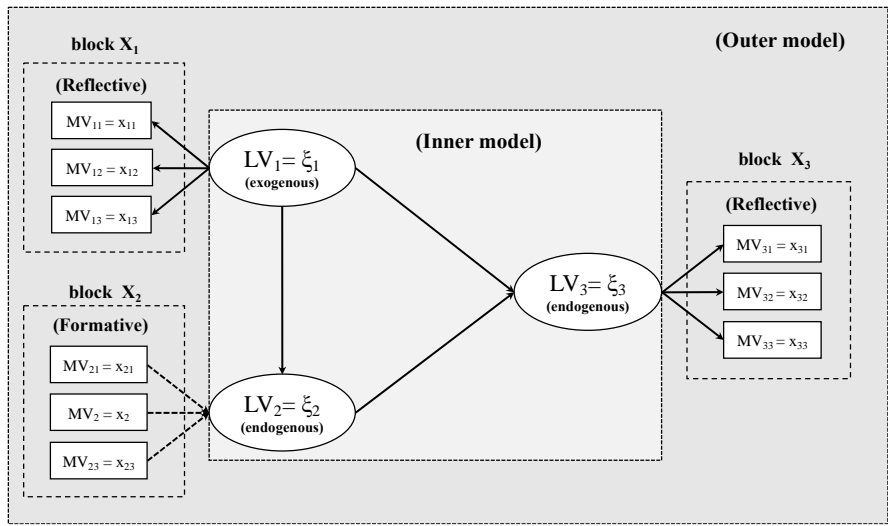

Figure 4. Simple causality path model.

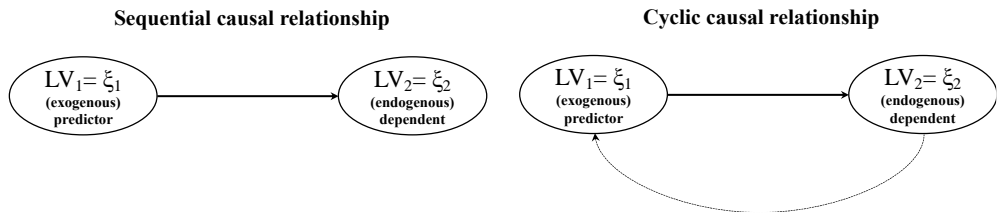

Figure 5. Two-step approach to estimating cyclic effects.

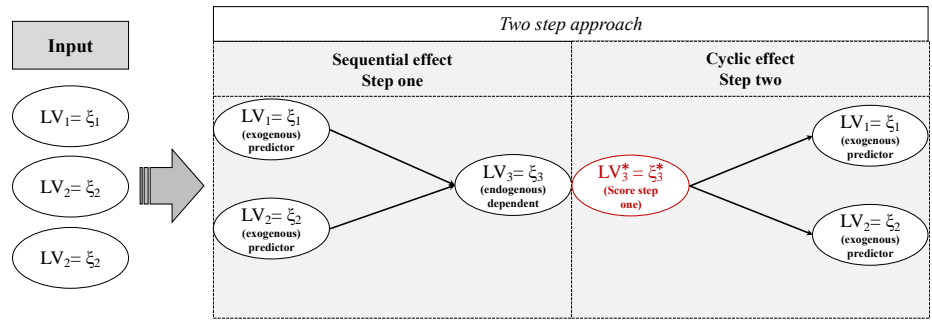

Figure 6. Two-step approach to estimating cyclic effects in internet appropriation

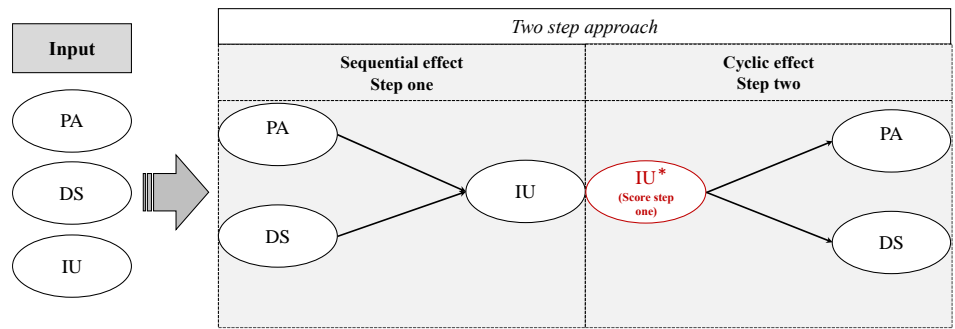

Figure 7. Sample characteristics.

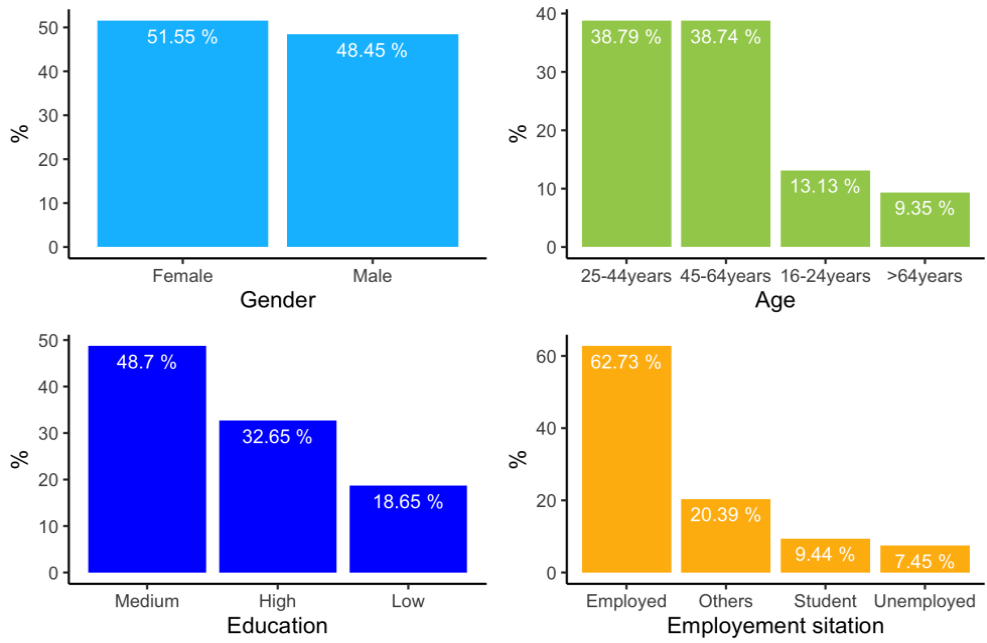

Figure 8. MCA factorial map of physical access intensity.

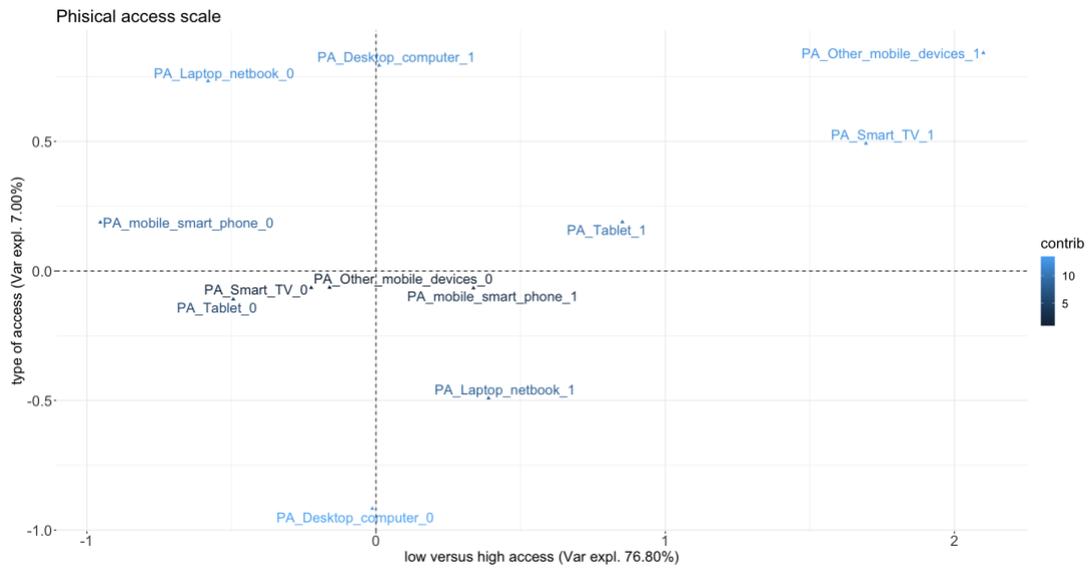

Figure 9. Sequential model estimation.

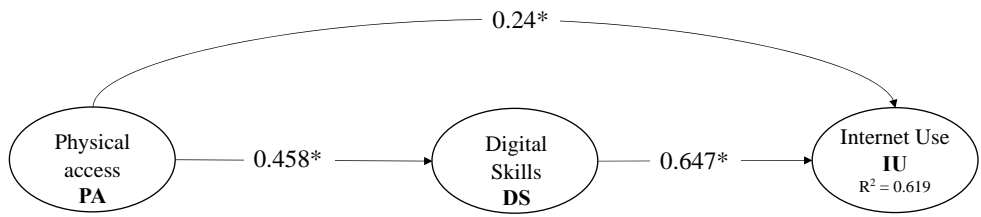

*Significant according to the CI at 95%

Figure 10. Cyclic model estimation.

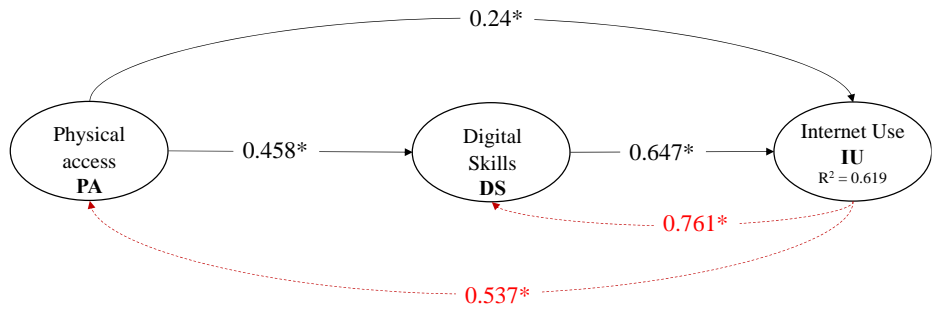

*Significant according to the CI at 95%